# Maximal Ratio Combining Diversity Analysis of Underwater Acoustic Communications Subject to $\kappa$–$\mu$ Shadowed Fading Channels

Ehab Salahat, *Student Member, IEEE*, Ali Hakam, *Student Member, IEEE*
{ehab.salahat;ali.hakam.a}@ieee.org

*Abstract*—In this paper, a novel unified analytical expression for average bit error rates (ABER) and average channel capacity (ACC) is presented for the $\kappa$–$\mu$ shadowed fading model. This model has recently shown to be suitable for underwater acoustic wireless channel modeling using the measurements conducted by the Naval Research Laboratory [1], and is not so well covered in the public literature. Deploying the maximal ratio combining diversity (MRC) receiver, a new simple analytical expression for the probability density function (PDF) of the receiver's output signal-to-noise ratio (SNR) is presented. Based on this new PDF, a novel unified ABER and ACC expression is derived. Moreover, to generalize the ABER analysis, the additive white generalized Gaussian noise model is assumed, which models different noise environments. The new unified expression is accurate, simple and generic, and is suitable for MRC analysis in this generalized shadowed fading and noise environment. Numerical techniques and published results from the literature prove the accuracy of our analysis and the generality of the derived unified expression.

*Keywords—Underwater Acoustic Channel; Additive White Generalized Gaussian Noise; κ-μ shadowed; Rician Shadowed.*

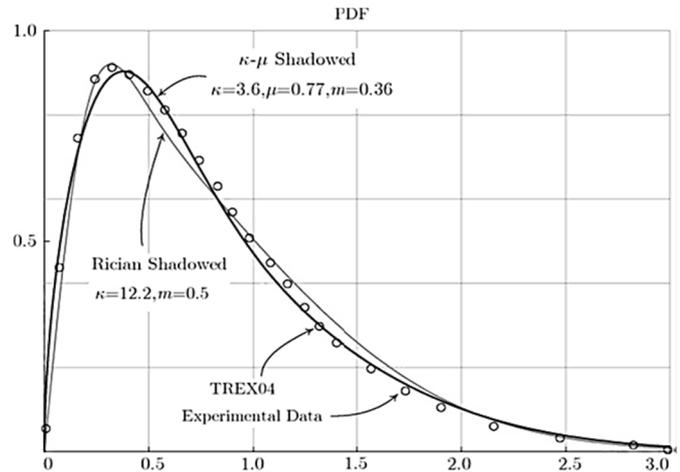

Fig. 1: Probability density function fitting plots for an UAC channel measured in the TREX04 experiment (see [1]).

## I. INTRODUCTION

WIRELESS COMMUNICATIONS is one of the most active and promising engineering research areas. It has a wide spectrum of limitless potential applications. Wireless communications systems are deployed in outdoor, indoor and enclosed (e.g. airplanes) environments, and recently in aqueous ones [2] [3]. Underwater acoustic communications systems (UAC) have attracted the interest of the research community with many foreseen futuristic and applications, e.g. ocean exploration, sea animals' observation and monitoring, military surveillance, unmanned underwater vehicles and submarines communications, etc. They have also evolved from analog to coherent digital systems [4].

Underwater acoustic communications (UAC) is, however, recognized to be one of the most difficult communication types in use today due to waves, internal turbulence, surface and seabed reflections, high signal attenuations, and other small-scale phenomena contributing to signal variations [5]. Statistical models are commonly used to characterize on the effects that degrade the signal during its propagation, namely shadowing and multipath fading. The awareness to develop standardized statistical distributions for UACs continues to increase; however, still addressed to a limited extent. Such statistical models will provide tools for predicting the oceanic system's performance prior to its deployment, and are hence essential for aqueous research [6].

The Rician shadowed fading model, proposed in [7], is a shadowed fading distribution that assumes Rician distribution for its line-of-sight component and Nakagami-*m* shadowing. This model was effectively used to model satellite channels as well as UACs [8]. The Rician shadowed fading distribution is a generalized model that includes many fading distributions as special cases. This model was investigated and analyzed using maximal ratio combining in [9].

The $\kappa$–$\mu$ shadowed fading model, introduced very recently in [1], is a new shadowed fading model that assumes Nakagami-*m* shadowing and $\kappa$–$\mu$ fading [10]. This shadowed fading model has a strong physical foundation and good analytical properties that allow it to be easily manipulated. It encloses the $\kappa$–$\mu$, $\eta - \mu$, Nakagami-*m*, Rician, Rician shadowed and the one-side Gaussian fading models as special cases [11]. This generalized model was shown to provide rather better fitting for the UAC measurements conducted by the Naval Research Laboratory, as compared to the Rician shadowed [1] [12], see Fig. 1. The new generalized shadowed fading model has analytically traceable statistical expressions (given in closed-form) with remarkable flexibility.

Driven by the above discussion, the goal of this research is to provide a rigorous analysis for $\kappa$–$\mu$ shadowed fading model that fits underwater acoustic propagation and other wireless fading scenarios. In parallel, we also aim to develop efficient and unified analytical expression that models the UACs in different noise and fading environments. The additive white generalized Gaussian noise (AWGGN) is assumed in this study. This noise model includes many other noise models (e.g. the additive white Gaussian noise (AWGN), the gamma, the Laplacian, the impulsive noise, and other noise models) as particular cases. Versatile closed-form mathematical models for the probability density function (PDF) and cumulative density function (CDF) for the combiner output signal-to-noise ratio, with the assumption of independent and identically distributed (i.i.d.) fading channels, are derived. We then adopt the approximation in [3] of the generalized $Q$–function and $\log_2(1+\gamma)$ in our derivations to present a novel unified and generic expression for the ABER and ACC to analyze wireless UACs. It should also be noted that the mathematical analysis can be directly extended to other wireless environments, but the emphasis in this paper is given to UAC.

The remaining part of this paper is structured as follows. In section II, the $\kappa$–$\mu$ shadowed fading model is visited, and the new simplified PDF and CDF expressions for the MRC output SNR are derived, with a quick revision to the generalized $Q$–function and its approximation. Next, the new unified ACC and ABER for MRC in AWGGN and $\kappa$–$\mu$ shadowed fading environment is presented in section III. Following to that, in section IV, the analysis and the results of the many test cases are presented, which are also corroborated by the numerical evaluation and compared to the published results from the technical literature. Finally, the paper contributions are then summarized in section V.

## II. MRC Statistical Derivations

### A. $\kappa - \mu$ Shadowed Fading Model – An Overview

The $\kappa$–$\mu$ shadowed fading model is a generalization of the physical $\kappa$–$\mu$ model [2] [10] [13]. In a non-homogeneous wireless propagation, the multipath waves are assumed to have scattered waves with the same power and an arbitrary power for the dominant component. Unlike the $\kappa$–$\mu$ model that adopts the assumption of having a deterministic dominant component within each multipath cluster, the $\kappa$–$\mu$ shadowed fading assumes that dominant components are subjected to shadowing.

Since the Rice model is a special case of the $\kappa$–$\mu$ distribution [10], a natural generalization of Rician shadowed model the $\kappa$–$\mu$ shadowed fading distribution [1]. It basically assumes that the LOS components are subjected to shadowing that follows Nakagami-$m$. Based on this assumption, it was in [1] that the PDF and CDF of the $\kappa$–$\mu$ shadowed fading are given respectively as:

$$f_\gamma(\gamma) = \frac{\mu^\mu m^m (1+\kappa)^\mu}{\Gamma(\mu)(\mu\kappa+m)^m \tilde\gamma^\mu} \gamma^{\mu-1} e^{-\left[\frac{\mu(1+\kappa)}{\tilde\gamma}\right]\gamma} {}_1F_1\left(m;\mu;\left[\frac{\mu^2\kappa(1+\kappa)}{[\mu\kappa+m]\tilde\gamma}\right]\gamma\right), \quad (1)$$

$$F_\gamma(\gamma) = \frac{\mu^{\mu-1} m^m (1+\kappa)^\mu \gamma^\mu}{\Gamma(\mu)(\mu\kappa+m)^m \tilde\gamma^\mu} \Phi_2\left(\mu-m, m; \mu+1; \frac{-\mu(1+\kappa)\gamma}{\tilde\gamma}, \frac{-\mu m(1+\kappa)\gamma^2}{\tilde\gamma(\mu\kappa+m)}\right). \quad (2)$$

In (1) and (2), $\kappa$ account for the ratio between the total power of the dominant components and the total power of the scattered waves, whereas $\mu$ and $m$ models the number of the multipath clusters and Nakagami-$m$ shadowing, respectively [1] [10], and the functions ${}_1F_1(\cdot)$ and $\Phi_2(\cdot)$, defined in [14], denote the confluent hypergeometric function and bivariate confluent hypergeometric function, respectively. The model includes the $\kappa$–$\mu$, $\eta-\mu$, Nakagami-$m$, Rician shadowed, Rician, Hoyt, Rayleigh and one-sided Gaussian as special cases. These distributions can be obtained from (1) as shown in and summarized in table I for brevity [1] [15].

TABLE I: PARTICULAR CASES OF THE $\kappa$–$\mu$ SHADOWED FADING [1] [15].

| Fading Distribution | $\mu$ | $\kappa$ | $m$ |
|---|---|---|---|
| $\kappa - \mu$ | $\mu$ | $\kappa$ | $m \to \infty$ |
| $\eta - \mu$ | $2\mu$ | $0.5\eta^{-1}(1-\eta)$ | $m = \mu$ |
| Rician shadowed | $\mu = 1$ | $\kappa = K$ | $m = m$ |
| Nakagami-$q$ (Hoyt) | 1 | $0.5q^{-2}(1-q^2)$ | $m = 0.5$ |
| Rician | $\mu = 1$ | $\kappa = K$ | $m \to \infty$ |
| Nakagami-$m$ | $\mu = m$ | $\kappa \to 0$ | $m \to \infty$ |
| Rayleigh | $\mu = 1$ | $\kappa \to 0$ | $m \to \infty$ |
| One-Sided Gaussian | $\mu = 0.5$ | $\kappa \to 0$ | $m \to \infty$ |

### B. Maximum Ratio Combiner Analysis

The maximal ratio combiner (MRC) receiver is one of the most studied receiver combiner designs. It generally provides convenient analytical traceability, allowing easy mathematical manipulations (see for example [2] [16]). We start this section by introducing the following theorem.

***Theorem:*** The PDF of the corresponding output SNR of the MRC receiver with the assumption of $L$ independently and identically distributed (i.i.d.) shadowed $\kappa$–$\mu$ fading channels is given by:

$$f_{\gamma_{MRC}}(\gamma) = \frac{\tilde\mu^{\tilde\mu}\tilde m^{\tilde m}(1+\kappa)^{\tilde\mu}}{\Gamma(\tilde\mu)(\tilde\mu\kappa+\tilde m)^{\tilde m}\eta^{\tilde\mu}} \gamma^{\tilde\mu-1} e^{-\left[\frac{\tilde\mu(1+\kappa)}{\eta}\right]\gamma} {}_1F_1\left(\tilde m; \tilde\mu; \left[\frac{\tilde\mu^2\kappa(1+\kappa)}{[\tilde\mu\kappa+\tilde m]\eta}\right]\gamma\right), \quad (3)$$

where $\tilde\mu = L\mu$, $\tilde m = Lm$, $\eta = L\tilde\gamma$, and $L$ defines the number of diversity branches adopted in the MRC receiver design.

***Proof:*** it follows from the i.i.d. assumption that the output and instantaneous SNR of the MRC is given [2] [17] [18] as:

$$\gamma = \sum_{i=1}^{L} \gamma_i. \quad (4)$$

and as was reported in [1], the moment generating function (MGF) of any of the $\kappa - \mu$ shadowed fading by:

$$M_\gamma(s) = \frac{(-\mu)^\mu m^m (1+\kappa)^\mu}{\tilde\gamma^\mu (\mu\kappa+m)^m} \frac{\left[s - \frac{\mu(1+\kappa)}{\tilde\gamma}\right]^{m-\mu}}{\left[s - \frac{\mu(1+\kappa)}{\tilde\gamma}\frac{m}{\mu\kappa+m}\right]^m}, \quad (5)$$

and further adopting the assuming that $\tilde\gamma = \tilde\gamma_1 = \tilde\gamma_2 = \cdots = \tilde\gamma_L$ for all the $L$ branches, it follows that the MGF of the MRC output obtained as:

$$M_{\gamma_{MRC}}(s) = \prod_{i=1}^{L} M_\gamma(s)$$

$$= \frac{(-\mu)^{L\mu} m^{Lm}(1+\kappa)^{L\mu}}{\tilde\gamma^{L\mu}(\mu\kappa+m)^{Lm}} \frac{\left[s-\frac{\mu(1+\kappa)}{\tilde\gamma}\right]^{Lm-L\mu}}{\left[s-\frac{\mu(1+\kappa)}{\tilde\gamma}\frac{m}{\mu\kappa+m}\right]^{Lm}}. \quad (6)$$

If, for compactness, new parameters are introduced as $\tilde\mu = L\mu$, $\tilde m = Lm$, and $\eta = L\tilde\gamma$, one can write (6) as:

$$M_{MRC}(s) = \frac{(-\tilde\mu)^{\tilde\mu}\tilde m^{\tilde m}(1+\kappa)^{\tilde\mu}}{\eta^{\tilde\mu}(\tilde\mu\kappa+\tilde m)^{\tilde m}} \frac{\left[s-\frac{\tilde\mu(1+\kappa)}{\eta}\right]^{\tilde m-\tilde\mu}}{\left[s-\frac{\tilde\mu(1+\kappa)}{\eta}\frac{\tilde m}{\tilde\mu\kappa+\tilde m}\right]^{\tilde m}}. \quad (7)$$

By taking the inverse Laplace transform of (7), or equivalently by comparing (5) and (7), one can conclude that the PDF of the output SNR with $L$ branches MRC receiver is given by:

$$f_{\gamma_{MRC}}(\gamma) = \frac{\tilde{\mu}^{\tilde{\mu}} \tilde{m}^{\tilde{m}} (1+\kappa)^{\tilde{\mu}}}{\Gamma(\tilde{\mu})(\tilde{\mu}\kappa+\tilde{m})^{\tilde{m}} \eta^{\tilde{\mu}}} \gamma^{\tilde{\mu}-1} e^{-\left[\frac{\tilde{\mu}(1+\kappa)}{\eta}\right]\gamma} {}_1F_1\left(\tilde{m}; \tilde{\mu}; \left[\frac{\tilde{\mu}^2\kappa(1+\kappa)}{[\tilde{\mu}\kappa+\tilde{m}]\eta}\right]\gamma\right). (8)$$

This new PDF will be very useful, due to its simplicity as compared to eqn. (10) in [1], which utilizes the complicated confluent hypergeometric function [19]. For compactness, (8) is written as:

$$f_{\gamma_{MRC}}(\gamma) = \psi \gamma^{\tilde{\mu}-1} e^{-\beta\gamma} {}_1F_1(\tilde{m}; \tilde{\mu}; \zeta\gamma), \quad (9)$$

where $\psi = \frac{\tilde{\mu}^{\tilde{\mu}} \tilde{m}^{\tilde{m}} (1+\kappa)^{\tilde{\mu}}}{\Gamma(\tilde{\mu})(\tilde{\mu}\kappa+\tilde{m})^{\tilde{m}} \eta^{\tilde{\mu}}}$, $\beta = \left[\frac{\tilde{\mu}(1+\kappa)}{\eta}\right]$, and $\zeta = \left[\frac{\tilde{\mu}^2\kappa(1+\kappa)}{[\tilde{\mu}\kappa+\tilde{m}]\eta}\right]$.

***Corollary:*** it follows from (2) and (9) that the corresponding CDF of the MRC reception will be given as:

$$F_\gamma(\gamma) = \frac{\tilde{\mu}^{\tilde{\mu}-1} \tilde{m}^{\tilde{m}} (1+\kappa)^{\tilde{\mu}} \gamma^{\tilde{\mu}}}{\Gamma(\tilde{\mu})(\tilde{\mu}\kappa+\tilde{m})^{\tilde{m}} \eta^{\tilde{\mu}}} \Phi_2\left(\tilde{\mu}-\tilde{m}, \tilde{m}; \tilde{\mu}+1; \frac{-\tilde{\mu}(1+\kappa)\gamma}{\tilde{\eta}}, \frac{-\tilde{\mu}\tilde{m}(1+\kappa)\gamma^2}{\tilde{\eta}(\tilde{\mu}\kappa+\tilde{m})}\right), (10)$$

This new form of generalized PDF in (9) is generic and suitable for all the fading models presented in table I with $L$-branches MRC reception. In section III, (9) will be used in the derivations of the unified ABER and ACC expression.

### C. The Generalized Q–Function Approximation

The Gaussian $Q$-function (see [20] and the references therein) has been widely studied in the wireless communications literature. It especially used to evaluate the average error rates of a given modulation scheme in the additive white Gaussian noise (AWGN). However, there are some situations that are encountered where the AWGN assumption might be invalid, and the use of other noise models might seem more suitable. The additive white generalized Gaussian noise (AWGGN) is a generalized noise model that includes many other noise models as particular cases, such as the AWGN, Laplacian, gamma, impulsive and others. The generalized $Q$–function is generally used in the formulation of the AWGGN noise and is given as [2][3][21]:

$$Q_a(x) = \frac{a\Lambda_0^{2/a}}{2\Gamma(1/a)} \int_x^\infty e^{-\Lambda_0^a |u|^a} du$$
$$= \frac{\Lambda_0^{2/a-1}}{2\Gamma(1/a)} \Gamma(1/a, \Lambda_0^a |x|^a). \quad (11)$$

where $\Lambda_0 = \sqrt{\Gamma(3/a)/\Gamma(1/a)}$, and $\Gamma(\cdot)$ is the gamma function. A summary of the relation between the AWGGN and other noise models is given in table II.

To simplify the mathematical derivations that involve the (generalized) $Q$–function, it would be helpful to have a simple yet accurate approximation that represents it. Moreover, since the $Q_a(\sqrt{\cdot})$ in the important form for error rates performance analysis, the following approx. was proposed in [3] as:

$$Q_a(\sqrt{x}) \approx \sum_{i=1}^{4} \delta_i e^{-\sigma_i x}, \quad (12)$$

TABLE II: Relation Between $Q_a(x)$ And Special Noise Models.

| Noise Dist. | Impulsive | Gamma | Laplacian | Gaussian | Uniform |
|---|---|---|---|---|---|
| $a$ | 0.0 | 0.5 | 1.0 | 2.0 | $\infty$ |

TABLE III: Fitting Parameters of $Q_a(\sqrt{\cdot})$ Approximation

| $a$ | $\delta_1$ | $\delta_2$ | $\delta_3$ | $\delta_4$ | $\sigma_1$ | $\sigma_2$ | $\sigma_3$ | $\sigma_4$ |
|---|---|---|---|---|---|---|---|---|
| 0.5 | 44.920 | 126.460 | 389.400 | 96.54 | 0.130 | 2.311 | 12.52 | 0.629 |
| 1 | 0.068 | 0.202 | 0.182 | 0.255 | 0.217 | 2.185 | 0.657 | 12.640 |
| 1.5 | 0.065 | 0.149 | 0.136 | 0.125 | 0.341 | 0.712 | 10.57 | 1.945 |
| 2 | 0.099 | 0.157 | 0.124 | 0.119 | 1.981 | 0.534 | 0.852 | 10.268 |
| 2.5 | 0.126 | 1.104 | -1.125 | 0.442 | 9.395 | 0.833 | 0.994 | 1.292 |

where $\delta_i$ and $\sigma_i$ are fitting parameters that were obtained offline using nonlinear curve fitting. Illustrative fitting values for several values of $a$ are presented in table III [3].

## III. Performance Evaluation

### A. Approximate Bit Error Rate Analysis

It's widely known that the average error rates (bit/symbol error rate) over fading channels can be evaluated by averaging the bit error rate of the noise channel using the fading PDF [20]. It can be expressed analytically expressed as:

$$P_e = \int_0^\infty f_\gamma(\gamma) \{\mathcal{A} Q_a(\sqrt{\mathcal{B}\gamma})\} d\gamma, \quad (13)$$

where $f_\gamma(\gamma)$ is the fading PDF of the $\kappa$–$\mu$ shadowed fading model, $\mathcal{A} Q_a(\sqrt{\mathcal{B}\gamma})$ is the error performance in AWGGN, and $\mathcal{A}$ and $\mathcal{B}$ are given in table IV for many coherent modulation schemes.

TABLE IV: $\mathcal{A}$ AND $\mathcal{B}$ VALUES FOR DIFFERENT MODULATIONS

| Modulation Scheme | Average SER | $\mathcal{A}$ | $\mathcal{B}$ |
|---|---|---|---|
| BFSK | $= Q_a(\sqrt{\gamma})$ | 1 | 1 |
| BPSK | $= Q_a(\sqrt{2\gamma})$ | 1 | 2 |
| QPSK, 4-QAM | $\approx 2 Q_a(\sqrt{\gamma})$ | 2 | 1 |
| M-PAM | $\approx \frac{2(M-1)}{M} Q_a\left(\sqrt{\frac{6}{M^2-1}\gamma}\right)$ | $\frac{2(M-1)}{M}$ | $\frac{6}{M^2-1}$ |
| M-PSK | $\approx 2 Q_a\left(\sqrt{2\sin^2\left(\frac{\pi}{M}\right)\gamma}\right)$ | 2 | $2\sin^2\left(\frac{\pi}{M}\right)$ |
| Rectangular M-QAM | $\approx \frac{4(\sqrt{M}-1)}{\sqrt{M}} Q_a\left(\sqrt{\frac{3}{M-1}\gamma}\right)$ | $\frac{4(\sqrt{M}-1)}{\sqrt{M}}$ | $\frac{3}{M-1}$ |
| Non-Rectangular M-QAM | $\approx 4 Q_a\left(\sqrt{\frac{3}{M-1}\gamma}\right)$ | 4 | $\frac{3}{M-1}$ |

Using (9), (13) is expressed as:

$$P_e = \mathcal{A}\psi \int_0^\infty \gamma^{\tilde{\mu}-1} e^{-\beta\gamma} {}_1F_1(\tilde{m}; \tilde{\mu}; \zeta\gamma) Q_a(\sqrt{\mathcal{B}\gamma}) d\gamma. \quad (14)$$

Furthermore, using the generalized $Q_a(\sqrt{\cdot})$ in (12), (13) can consequently be represented as:

$$P_e = \mathcal{A}\psi \sum_{i=1}^{4} \delta_i \int_0^\infty \gamma^{\tilde{\mu}-1} e^{-[\beta+\sigma_i \mathcal{B}]\gamma} {}_1F_1(\tilde{m}; \tilde{\mu}; \zeta\gamma) d\gamma, \quad (15)$$

with simple variable transform, as $z = \zeta\gamma$, (15) is evaluated in a simple closed form as:

$$P_e = \sum_{i=1}^{4} \Psi \, {}_2F_1\left([\tilde{m}, \tilde{\mu}]; [\tilde{\mu}]; \tilde{\beta}^{-1}\right), \quad (16)$$

with $\tilde{\beta} = [\beta + \sigma_i \mathcal{B}]/\zeta$ and $\Psi = [\mathcal{A}\psi\delta_i \Gamma(\tilde{\mu})]/[\zeta^{\tilde{\mu}} \tilde{\beta}^{\tilde{\mu}}]$.

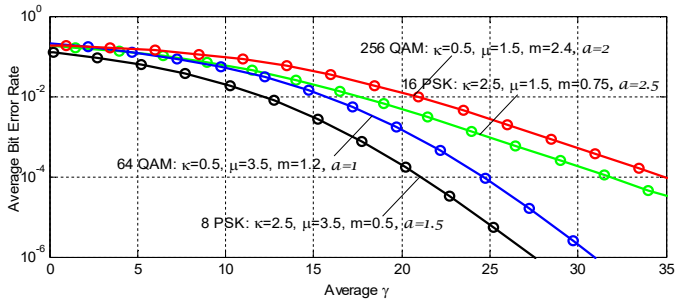

Fig. 2: ABER using MRC with $L=1$ in $\kappa$–$\mu$ shadowed fading.

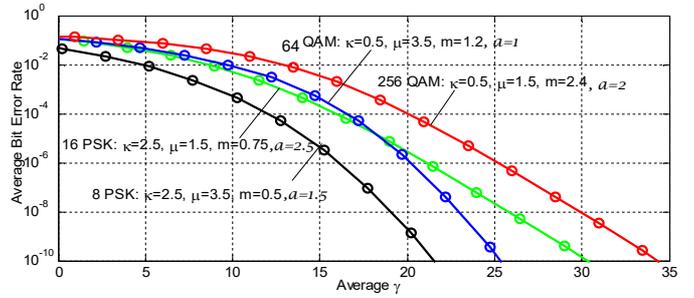

Fig. 3: ABER using MRC with $L=3$ in $\kappa$–$\mu$ shadowed fading.

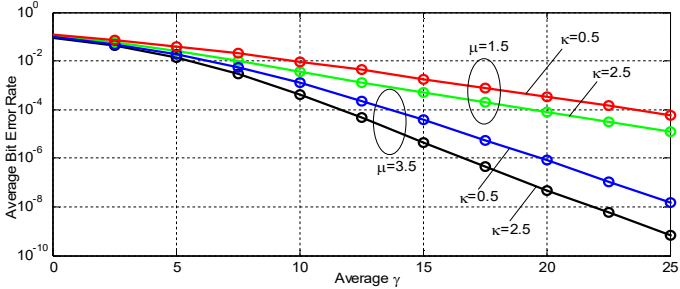

Fig. 4: ABER for BPSK using MRC with $L=1$ in $\kappa$–$\mu$ fading.

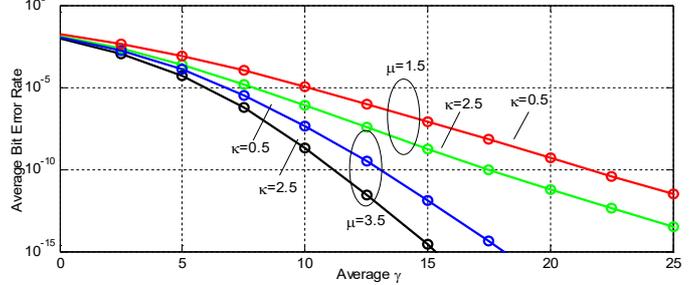

Fig. 5: ABER for BPSK using MRC with $L=3$ in $\kappa$–$\mu$ fading.

This new expression in (16) is simpler than that in [1] in eqn. (21). It also generalizes the analysis of MRC reception to the special cases of the $\kappa$–$\mu$ shadowed fading (such as the Rician shadowed, Nakagami-$m$, etc.) in different noise environments.

### B. Average Channel Capacity of AWGN Channel

The Channel capacity (given in bits/seconds) is defined as the maximum deliverable information rate over a transmission channel with arbitrarily and negligible error. The average channel capacity (a.k.a. ergodic capacity), can be attained by averaging the capacity of AWGN channels over the fading PDF [17]. This is represented as:

$$C = \int_0^\infty \log_2(1+\gamma) f_\gamma(\gamma) d\gamma. \quad (17)$$

C in this context is the normalized channel capacity, given in bits/second/Hz. Adopting a similar exponential approximation of $\log_2(1+\gamma)$ as the one provided in [3] which is also similar to the approx. of the $Q_a(\sqrt{\cdot})$ in (12), one has:

$$\log_2(1+\gamma) \approx \sum_{i=1}^4 \delta_i e^{-\sigma_i \gamma}, \quad (18)$$

then the average channel capacity (ACC) can be derived in exactly the same form as that of the ABER for this shadowed fading model (16) while adopting the new fitting values of $\delta$ =[9.331, -2.635, -4.032, -2.388] and $\sigma$=[0.000, 0.037, 0.004, 0.274] as was given in [3], Fig. 4. This new expression is appropriate for a reasonably large SNR range, which is the SNR range of interest in many wireless system scenarios that are studied in the literature.

### IV. RESULTS

In this section, the purpose is to validate the accuracy and capabilities of the derived unified ABER and ACC expression. This will be done by comparing the plots generated via the unified ABER and ACC expression derived in this paper versus the plots obtained using numerical evaluation. In addition, some scenarios will also be compared with results that are available in the technical literature for some of the particular cases of the $\kappa - \mu$ shadowed fading model.

Five test scenarios will be illustrated for the ABER here based on (16). The first test considers the general $\kappa$–$\mu$ shadowed fading in AWGGN, using different modulation schemes from Table IV of different constellation orders, namely 8-PSK, 16-PSK, 32-QAM and 256-QAM, using different values of model parameters and subject to different noise environments, specified by the value of $a$, without diversity (i.e. $L$=1). The ABER curves of this case are shown in Fig. 2. One can clearly see that numerically integrated results (solid lines) match the results of (16) denoted by the overlaid dots, confirming its accuracy for the different test scenarios presented. The second test case, presented in Fig. 3, uses the same previous values of parameters subject to the same noise environments, but deploying MRC reception with three diversity branches ($L$=3). It's obvious that (16) gives very accurate results as the dots overlay the curves of the numerical integration.

As a third test case, we consider the special case of $\kappa$–$\mu$ fading using BPSK in AWGN and selected the $\kappa$ and $\mu$ values to match those found in [2], to compare our generated results, shown in Fig. 4, with those in the technical literature. One can see a clear match between the ABER curves in Fig. 4 and those shown in [2]. Fig. 5 represents the same previous test case with the exception of increasing $L$ to 3 instead. Again, and as expected from using more diversity antenna, the performance has improved, and the figure shows an excellent agreement between numerical and analytical plots.

As for the Average Channel Capacity (ACC), one test case is presented due to space limitations. The ACC of the $\kappa$–$\mu$

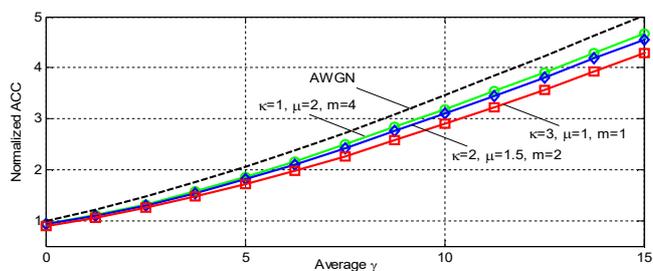

Fig. 6: ACC in different κ–μ shadowed fading scenarios.

shadowed fading is considered, and the generated results are shown in Fig. 6. Following on the discussion presented in section III.B, the ACC expression is exactly the same as that in (16) with changing the fitting parameters to the new ones specified there. The results of (16) agree with the curves of the numerical integration (solid lines) for the different tested cases. This proves that (16) is indeed unified for ABER and ACC analysis.

## V. Conclusion

In this paper, we presented novel performance analysis of the new κ–μ shadowed fading model, which was shown to be suitable for underwater acoustic communications channel modeling. Deploying the MRC reception, we derived a new PDF for the MRC output SNR. We also utilized the generalized $Q$-function approximation used to model the AWGGN in our ABER analysis. Consequently, a novel unified ABER and ACC expression for the analysis of κ–μ shadowed fading in AWGGN conditions using $L$-branch MRC receiver. The unified expression was shown to be very accurate with a wide spectrum of promising applications due the generality of this expression.